\documentclass[aps,prb,twocolumn,showpacs,floatfix]{revtex4}
\usepackage{graphicx}
\usepackage{latexsym}
\usepackage{amsmath}
\usepackage{graphics}
\usepackage{amssymb}
\usepackage{layout}
\usepackage{amsfonts,epsfig}
\newcommand{\ket}[1]{\left|#1\right>}
\newcommand{\bra}[1]{\left< #1 \right|}

\begin{document}
\title{Quasiparticle decay rate of Josephson charge qubit oscillations}

\author{Roman Lutchyn$^{1}$, Leonid Glazman$^{1}$, and Anatoly Larkin$^{1,2}$}

\affiliation{$^{1}$ W.I.\ Fine Theoretical Physics Institute,
University
of Minnesota, Minneapolis, Minnesota 55455, USA \\
$^{2}$ L. D. Landau Institute for Theoretical Physics, Moscow,
117940, Russia}

\date{\today }
\begin{abstract}
    We analyze the decay of quantum oscillations in a charge qubit
  consisting of a Cooper pair box connected by a Josephson junction to
  a finite-size superconductor. We concentrate on the contribution of
  quasiparticles in the superconductors to the decay rate. Passing of a quasiparticle through the Josephson junction leads to
  the escape of the qubit out of its Hilbert space, and thus
  determines the decay rate of quantum oscillations. We
  find the temperature dependence of the quasiparticle contribution to
  the decay rate for open and isolated systems. The former case is realized if a normal-state
  trap is included in the circuit, or if just one vortex resides in
  the qubit; we find exponential suppression of the rate,
  $\Gamma\propto\exp\left(-\Delta/T\right)$, at low temperatures (here, $\Delta$
  is the superconducting gap). In a superconducting qubit isolated from
  the environment $\Gamma\propto\exp\left(-2\Delta/T\right)$ if the number of electrons is even,
  while for an odd number of electrons the decay rate
  remains finite in the limit $T\to 0$. We estimate $\Gamma$ for
  realistic parameters of a qubit.
\end{abstract}
\pacs{03.65.Yz, 03.67.Lx, 74.50.+r, 85.25.Cp}

\maketitle

\section{Introduction}
Recent experiments with superconducting qubits show promising
results, allowing one to observe hundreds of qubit charge or phase
oscillations~\cite{Nakamura, Wallraff, Vion, Chiorescu}. A
superconducting qubit is essentially a controllable quantum
two-level system and can be realized using phase or charge degrees
of freedom.  Most of the charge qubits use a Cooper-pair box
(CPB), a small mesoscopic island connected to a large
superconducting reservoir via two Josephson junctions. A device
with a large superconducting gap, $\Delta> E_c \gtrsim E_J \gg T$,
can be controlled with the gate voltage and magnetic flux, and has
only one discrete degree of freedom: number of Cooper pairs in the
box. (Here, $E_c$ is the charging energy of the island, and $E_J$
is an effective Josephson energy of its junctions with the
reservoir; $E_J$ depends on the flux.) At the degeneracy point the
qubit evolves coherently between the ground and excited states,
$\ket{-}$ and $\ket{+}$, which are represented by the
superposition of 0 or 1 excess Cooper pairs in the box. The
frequency of these oscillations is determined by the Josephson
energy.

The main difficulty in technological realization of a charge-based
superconducting qubit is due to the decoherence present in it. The
mechanisms of the decoherence are currently unknown and can be
attributed to phonons, two-level systems in insulating barrier,
thermally excited quasiparticles~\cite{Blais, Devoret}. In this
paper we concentrate on the contribution of the quasiparticles to
the decay rate. Quasiparticles having a continuous spectrum are
inherently present in any superconducting device and set a
fundamental constraint on the coherence time.  Quasiparticle
``poisoning'', first investigated in the context of charge-parity
effects in mesoscopic superconductors~\cite{ Tinkham, Hekking,
Lafarge}, manifests itself also in the experiments with charge
qubits~\cite{Aumentado, Turek, Guillaume, Lehnert, Mannik}.  It
was reported that even at low temperatures ($\sim$10-50mK)
quasiparticles are present in the CPB.  If this is the case,
Hilbert space of the CPB expands, and the qubit is no longer a
simple two-level system. The transient presence of a quasiparticle
in the CPB detunes the qubit from the resonant state of Cooper
pair tunneling and affects coherent oscillations. In this article
we build a quantitative theory of the quasiparticle effect on the
charge qubit oscillations.

We consider two regimes which can be realized experimentally: open
system corresponding to fixed chemical potential in the reservoir,
and isolated system corresponding to fixed number of electrons in
the qubit (see Fig.1). The former case allows for a change of the
total number of electrons in superconducting parts of the system,
and is experimentally realized if a quasiparticle trap, e.g.
normal-state part or a vortex, is included in the circuit. The
latter case corresponds to a superconducting qubit isolated from
the normal-metal environment. Both cases may be relevant in the
context of the cavity quantum electrodynamics experiments where
the state of the qubit is determined using photon degrees of
freedom.

The paper is organized as follows. We begin in Sec. II with a
brief overview of the charge qubit before the discussion of the
quasiparticle effect on the charge qubit in an open system. In
Sec. III we consider the opposite case of the isolated qubit.
Finally, in Sec. IV we present simplified results for the
quasiparticle contribution to the decoherence rate for the
mentioned experimental realizations and discuss how to decrease
it.
\section{States of the qubit in an open system}
Dynamics of the superconducting charge qubit is usually described
by an effective Hamiltonian
\begin{equation}\label{Hqubit}
H_{\!eff}=E_c(N-N_g)^2+H_J,
\end{equation}
where $E_c$ is charging energy of the box, $N$ is dimensionless
charge of the CPB, and $N_g$ is dimensionless gate voltage. The
Hamiltonian corresponding to tunneling of Cooper pairs, $H_J$, is
defined as $H_J=-\frac{E_J}{2}\left(\ket{N+2}\bra{N}+h.c.\right)$,
where $E_J$ is effective Josephson energy. In the regime where
superconducting gap is the largest energy scale in the system
$\Delta> E_c\gtrsim E_J \gg T$, quasiparticles are usually
neglected, and the dynamics of the system is described by the
above Hamiltonian, where there is only one discrete degree of
freedom - excess number of Cooper pairs in the box. At the
operating point, when the dimensionless gate voltage is tuned to
be equal to one, $N_g=1$, there is degeneracy with respect to
charging energy between the charge states $\ket{N}$ and
$\ket{N+2}$. This degeneracy is lifted by the Josephson energy,
and the states of a qubit are described by the symmetric and
antisymmetric superposition of the charge states,
$\ket{-}=\frac{\ket{N+2}+\ket{N}}{\sqrt{2}}$ and
$\ket{+}=\frac{\ket{N+2}-\ket{N}}{\sqrt{2}}$ with corresponding
energies:
\begin{eqnarray}\label{omega}
\omega_{-}=E_c-\frac{E_J}{2} \mbox{ and }
\omega_{+}=E_c+\frac{E_J}{2}.
\end{eqnarray}
Other charge states have much higher energy, and effectively the
Cooper pair box reduces to a two-level system. Once a qubit is
excited, quantum oscillations between  states $\ket{-}$ and
$\ket{+}$ emerge, and the frequency of these oscillations is
determined by the Josephson energy $E_J$~\cite{Shnirman}.
\begin{figure}
\centering
\includegraphics[width=2.7in]{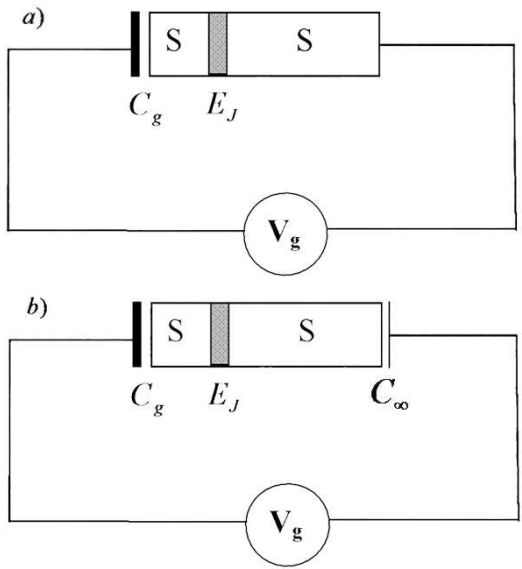}
\caption{ Schematic picture of the charge qubit in different
experimental regimes: a) open system, b) isolated system. The left
superconducting mesoscopic island is the Cooper pair box connected
via a tunable Josephson junction to the large superconducting
reservoir (right). Gate bias is applied through the capacitance
$C_g$ (assuming that $C_{\infty}\gg C_g$).}
\end{figure}
The appearance of a quasiparticle with a continuum spectrum
provides a channel for relaxation of the qubit. Since
quasiparticles are inherently present in any superconducting
system, their contribution to the decay rate is intrinsic.

\subsection{Thermodynamic properties of a qubit in an open system}
At a finite temperature, density of quasiparticles is
exponentially small, $n\propto \exp\left(-\Delta/T\right)$, but
the number of quasiparticles in the grain can be of the order of
one. It is important to point out that even one unpaired electron
can affect the qubit performance. In order to estimate the number
of quasiparticles in the grains, one has to account for the huge
statistical weight of the states with a single quasiparticle,
proportional to the volume of the grain, $N \simeq
\sqrt{2\pi\Delta_b T}\nu_b V_b
 \exp\left(-\Delta_b/T\right)$, where $\nu_b$ is normal density of
 states(per volume), $V_b$ volume of the grain. For an isolated dot, the characteristic temperature~\cite{Matveev}
at which quasiparticles appear is ($k_B=1$)
\begin{equation}\label{tstar}
T_{b}^*=\frac{\Delta_b}{\ln(\Delta_b/\delta_b)}.
\end{equation}
where $\delta_{b}=\frac{1}{\sqrt{2\pi} \nu_b V_b}$ is the mean level
spacing in the box, and $\Delta_b$ is superconducting gap energy in
the box. The appearance of a quasiparticle in the qubit occurs at
lower temperature $\widetilde{T}_{b}^*$ due to the finite charging
energy:
\begin{equation}
\widetilde{T}_{b}^*=\frac{\Delta_b-E_c+E_J/2}{\log(\Delta_b/\delta_{b})}=T_{b}^*\left(1-\frac{E_c-E_J/2}{\Delta_b}\right).
\end{equation}
If $T \ll \widetilde{T}_{b}^*$, states with odd number of
electrons in the box are statistically rare. The probability of
finding the qubit in a ``good" state (not poisoned by
quasiparticles) is important for qubit preparation and is
determined by thermodynamics. However, the qubit coherence time is
controlled by kinetics, which we study in the next section.

\subsection{Quasiparticle decay rate in the open system}

If an unpaired electron tunnels into the CPB, it tunes the qubit
away from the resonant state of Cooper pair tunneling, which leads
to the decay of quantum oscillations. At the operating point
$N_g=1$ it is energetically favorable for a quasiparticle to
tunnel to the CPB, since charging energy is gained in such
process. Assuming that initially the qubit was prepared in the
state with no quasiparticles in the box, the lifetime of the qubit
is determined by the time of quasiparticle tunneling to the CPB.
In order to estimate this time we use the following Hamiltonian:
\begin{equation}\label{1}
H=H_0^{'} + H_{T},
\end{equation}
\begin{equation}
 H_0^{'} = H^L_{BCS}+H^R_{BCS}+E_c(Q/e-N_g)^2, \nonumber
\end{equation}
where $H^L_{BCS}$, $H^R_{BCS}$ are BCS Hamiltonians of the box and
superconducting reservoir (see Fig. 1), and $Q$ denotes the charge
in the box. The tunneling Hamiltonian between two electrodes,
$H_T$, is defined as
\begin{equation}\label{HT}
H_{T}=\sum_{kp\sigma}(t_{kp}c_{k,\sigma}^{\dag}c_{p,\sigma}+\emph{H.c.})
\end{equation}
where $t_{kp}$ is tunnelling matrix element, $c_{k,\sigma}$,
$c_{p,\sigma}$ are the annihilation operators for an electron in
the state $k,\sigma$ in the CPB and state $p,\sigma$ in the
superconducting reservoir.

 We now consider the lifetime of the qubit in the open system,
 allowing for a change in the number of electrons in the superconducting reservoir.
 Assuming that the qubit was prepared in the initial state without
 a quasiparticle in the box, the time of its coherent evolution is limited
 by the rate of quasiparticle tunneling to the CPB. In order to
 calculate the lifetime of the qubit, we have to
 distinguish between tunneling of Cooper pairs and quasiparticles.
 To do this, we write the Hamiltonian $H$ in Eq.~(\ref{1}) in the following
 form:
 \begin{equation}\label{H0}
H=H_0 + H_{1},
 \end{equation}
where $H_0=H_0^{'}+H_J$, and $H_1=H_T-H_J$. $H_J$ is second order
in tunneling amplitude
\begin{equation}
H_J=\ket{N}\bra{N} H_T \frac{1}{E-H_0^{'}}H_T
\ket{N\!+2}\bra{N\!+2} +H.c.
\end{equation}
The matrix element $\bra{N} H_T \frac{1}{E-H_0^{'}}H_T \ket{N+2}$
is proportional to effective Josephson energy $E_J$, and $H_T$ is
defined in Eq.~(\ref{HT}). Without quasiparticles Hamiltonian
$H_0$ reduces to Eq.~(\ref{Hqubit}).

The quasiparticle tunneling rate is found using Fermi's golden
rule
 and averaging over initial configuration with the appropriate
 density matrix ($\hbar=1$)
\begin{equation}\label{Golden}
\Gamma=2\pi\sum_{i,f}|\bra{f}H_{1}\ket{i}|^2\delta(E_{f}-E_{i})\rho(\beta
H_0).
\end{equation}
Here, $\rho(\beta H_0)$ is the density matrix for the initial
state of the system. The perturbation Hamiltonian accounts for
quasiparticle tunneling only,
$\bra{f}H_{1}\ket{i}=\bra{f}H_{T}\ket{i}$. Thus,
Eq.~(\ref{Golden}) takes into account Cooper pair tunneling
exactly while treating quasiparticle tunneling perturbatively. At
the operating point, $N_g=1$, initial state of qubit is defined by
the superposition of 0 and 1 excess Cooper pairs in the box,
$\ket{\pm}=\frac{\ket{N}\mp\ket{N+2}}{\sqrt{2}}$; the final state
$\ket{f}=\ket{N+1}$ is the state with odd number of electrons in
the CPB, corresponding to charge 1$e$. There are two mechanisms
that contribute to the rate of the process $\ket{\pm} \rightarrow
\ket{N+1}$: (1) a quasiparticle tunnels from the superconducting
reservoir to the CPB, and (2) a Cooper pair in the box breaks into
two quasiparticles, and then one quasiparticle tunnels out into
the reservoir. Two corresponding contributions to the total
tunneling rate  are
\begin{equation}\label{total}
\Gamma_{\pm}=\Gamma_{1\pm}+\Gamma_{2\pm},
\end{equation}
\begin{eqnarray}\label{total1}
\Gamma_{1\pm}&=&2\pi\!\sum_{n,p_j,k_i}|\bra{N+1, k,\{p \}_{n-1}}
H_{T}\ket{\pm,\{p\}_{n}}|^2 \nonumber\\
 &\times&\delta(E_{k}-E_{p}-\omega_{\pm})\rho(\beta H_0),
\end{eqnarray}
\begin{eqnarray}\label{total2}
\Gamma_{2\pm}&=&2\pi\!\sum_{n,p_j,k_i}|\bra{N+1,
p,\{k\}_{2n-1}}H_{T}\ket{\pm,\{k\}_{2n}}|^2 \nonumber\\
 &\times&\delta(E_{p}-E_{k}-\omega_{\pm})\rho(\beta H_0),
\end{eqnarray}
 where $\Gamma_{\pm}$ is decay rate for the excited $\ket{+}$ or ground
 state $\ket{-}$
 of the qubit, and $\omega_{\pm}$ is defined in Eq.~(\ref{omega}). State $\ket{+,\{p\}_n}$, for example,
 denotes the excited state of the qubit with $n$ quasiparticles in the reservoir with energies $E_{p}=\sqrt{\varepsilon_{p}^2+\Delta_{r}^2}$
\begin{equation}
\ket{+,\{p\}_{n}}=\ket{+}\otimes \ket{p_1,\ldots,p_j,\ldots,
p_{n}}.
\end{equation}
The state $\ket{+,\{k\}_{2n}}$ denotes the excited state of the
qubit with $n$ broken Cooper pairs in the box,
 leading to the appearance of $2n$ quasiparticles with energies $E_k=\sqrt{\varepsilon_{k}^2+\Delta_{b}^2}$
\begin{equation}
\ket{+,\{k\}_{2n}}=\ket{+,k_1,\ldots,k_j,\ldots, k_{2n}}.
\end{equation}

In the following, we concentrate on the decay rate of the qubit
excited state in the open system, \textit{i.e.}, evaluate
$\Gamma_{+}^{\rm{op}}$.
 In order to calculate this decay rate we take into account
 one-electron processes in the lowest order in quasiparticle
 density. We assume that in the first contribution to $\Gamma_{+}^{\rm{op}}$,
 Eq.~(\ref{total1}), all quasiparticles are in the
 reservoir and one of them is tunneling into an unoccupied state of the CPB; in the
 second contribution, Eq.~(\ref{total2}), all quasiparticles are in the box and one of them
 is tunneling out into an unoccupied state of the reservoir. Keeping this in mind, the density matrix for first process can be
 reduced by tracing out irrelevant degrees of freedom: $\rho_{\rm{op}}(\beta H_0)=Tr_{\{k\}}\rho(\beta H_0)$, and Eq.~(\ref{total1}) becomes
\begin{eqnarray}\label{15}
\Gamma_{1}^{\rm{op}}&=&\pi\!\sum_{n,p_j,k}|\bra{N+1, k,\{p
\}_{n-1}}
H_{T}\ket{N,\{p\}_{n}}|^2 \nonumber\\
 &\times&\delta(E_{k}-E_{p}-\omega_{\pm})\rho_{\rm{op}}(\beta H_0).
\end{eqnarray}
Taking into account that only one quasiparticle is transferred
through the junction by the action of Hamiltonian $H_T$, and
performing the sum
 over momenta in Eq.~(\ref{15}), the contribution of the first mechanism is simplified to
\begin{eqnarray}\label{Gamma1}
\Gamma_{1}^{\rm{op}}&=&\!\pi\sum_{p_{1},k}|\bra{N\!+1,k}H_{T}\ket{N,p_{1}}|^2\delta(E_{k}-E_{p_{1}}-\omega_{+})
\nonumber \\
&\times& \exp\left(-\frac{E_{p_{1}}}{T}\right),
\end{eqnarray}
where the exponential factor is the low temperature
($T\ll\Delta_r$) approximation of the Fermi function. The matrix
element $\bra{N+1,k}H_{T}\ket{N,p}$ can be calculated using the
particle conserving Bogoliubov transformation~\cite{Schrieffer}
and is equal to $\bra{N\!+1,k}H_{T}\ket{N,p}=2(t_{pk}u_p u_k -
t_{kp}v_p v_k)$, where $u_p$, $v_p$ are coherence factors
\begin{eqnarray}
u_p^2=\frac{1}{2}\left(1+\frac{\varepsilon_{p}}{E_p}\right),
%\nonumber \\
%\nonumber \\
\mbox{ and }
v_p^2=\frac{1}{2}\left(1-\frac{\varepsilon_{p}}{E_p}\right).
\nonumber
\end{eqnarray}
By changing the sum to an integral in Eq.~(\ref{Gamma1}) and
integrating over $E_k$, we get the following expression for
$\Gamma_{1}^{\rm{op}}$:

\begin{eqnarray}\label{Gamma11}
\Gamma_{1}^{\rm{op}} &=&
\frac{g}{4\pi}\int_{\Delta_r}^{\infty}{dE_p}\frac
{\Theta(E_p+\omega_{+}-\Delta_b)}{\sqrt{((E_p+\omega_{+})^2-\Delta_b^2)(E_p^2-\Delta_r^2)}} \nonumber\\
%\nonumber \\
&\times&(E_p(E_p+\omega_{+})-\Delta_r
\Delta_b)\exp\left(-\frac{E_p}{T}\right),
\end{eqnarray}
where $\Theta(x)$ is the step function, $g=\frac{h}{e^2 R}$ is
dimensionless conductance. R is resistance of the tunnel junction
in the normal state
\begin{eqnarray}
%\nonumber \\
%\frac{1}{R}
R^{-1} =4\pi e^2\sum_{p,k}|t_{pk}|^2
\delta(\varepsilon_p)\delta(\varepsilon_k) \nonumber.
\end{eqnarray}
Assuming that mismatch between superconducting gap energies in the
box and reservoir is small, $\Delta_r-\Delta_b+\omega_{+}
> 0$, which corresponds to most charge qubit
experiments, expressions for $\Gamma_{1}^{\rm{op}}$ can be
simplified. The leading contribution to the decay rate at low
temperatures given by the first mechanism  is equal to
$\Gamma_{1}^{\rm{op}}=W(\omega_{+}, \Delta_r,\Delta_b)$
\begin{widetext}
\begin{eqnarray}\label{Gamma0}
W(\omega_{+},\Delta_r,\Delta_b)= \frac{g\sqrt{\Delta_r
}}{8\sqrt{2\pi}} \frac{(\Delta_r\!-\!\Delta_b\!+\!\omega_{+})
}{\sqrt{\Delta_r\!+\!\Delta_b\!+\!\omega_{+}
}}\exp{\!\left(\!-\frac{\Delta_r}{T}\right)}\!\left[\frac{\Delta_b}{\Delta_r}
U\left(\frac{3}{2},2,\!\frac{\Delta_r\!-\!\Delta_b\!+\!\omega_{+}}{T}\right)\!+\!2U\left(\frac{1}{2},2,\!\frac{\Delta_r\!-\!\Delta_b\!+\!\omega_{+}}{T}\right)\right],
\end{eqnarray}
\end{widetext}
where $U(a,b,z)$ is the confluent hypergeometric function. At low
temperature ($T\ll \Delta_r-\Delta_b+\omega_{+}$) asymptotic
result for $W(\omega_{+}, \Delta_r,\Delta_b)$ is simply given by
\begin{equation}
W(\omega_{+},\! \Delta_r,\Delta_b)\! \simeq\!
\frac{g\sqrt{\Delta_r T}}{4\sqrt{2\pi}}
\sqrt{\frac{\Delta_r\!-\!\Delta_b\!+\!\omega_{+}}{\Delta_r\!+\!\Delta_b\!+\!\omega_{+}
}}\exp\left(\!-\frac{\Delta_r}{T}\right).
\end{equation}
As expected, the decay rate due to the first mechanism is
exponentially suppressed due to the fact that it costs energy
$\Delta$ to bring a quasiparticle from the normal parts.

The contribution of the second mechanism given by
Eq.~(\ref{total2}) depends on the density matrix of the box. The
initial state of the qubit corresponds to the even-charge state in
the CPB. Statistical weight of the states with even number of
quasiparticles in the dot, $2, 4, 6,..., 2n$, is determined by the
density matrix $\rho_{2n}(\beta H_0))$,
\begin{equation}\label{rho}
\rho_{2n}(\beta H_0))=Tr_{\{p\}}\rho(\beta
H_0)=\frac{\exp\left(-\sum\limits_{j=2}^{2n}{\frac{E_{k_j}}{T}}\right)}{(2n)!
Z_{\rm{ev}}},
\end{equation}
where $Z_{\rm{ev}}=\cosh\left(z_b(T,\delta_b)\right)$ is the
partition function for the dot with even number of
electrons~\cite{Matveev}, and $z_{b}(T,\delta_{b})$ is
\begin{equation}
z_{b}(T,\delta_{b})=\sum_{k}\exp\left(-\frac{E_{k}}{T}\right)=
\sqrt{\frac{T}{\Delta_{b}}}\frac{\Delta_{b}}{\delta_{b}}\exp\left(-\frac{\Delta_{b}}{T}\right).
\end{equation}
According to Eq.~(\ref{total2}) and Eq.~(\ref{rho}), the
contribution to the decay rate due to the second mechanism is
obtained by summing over the states with even number of
quasiparticles with appropriate statistical weight
\begin{eqnarray}\label{2n}
\Gamma_{2}^{\rm{op}}&=& 2\pi\sum_{n,p,k_j}|\bra{N+1,p,\{k\}_{2n-1}}H_{T}\ket{+,\{k\}_{2n}}|^2 \nonumber\\
&\times& \delta(E_{p}-E_{k_1}-\omega_{+})2n\rho_{2n}(\beta H_0),
\end{eqnarray}
where, for example, $\bra{N+1,p,\{k\}_{2n-1}}$ is a state
corresponding to the charge on the box equal to 1$e$, $2n-1$
quasiparticles in the box, and 1 quasiparticle in the reservoir
\begin{equation}
\ket{N+1,p,\{k\}_{2n-1}}=\ket{N+1,k_1,\ldots,k_j,\ldots,
k_{2n-1}}\otimes \ket{p}.
\end{equation}
The additional factor of $2n$ in Eq.~(\ref{2n}) is the result of
the summation of $2n$ identical terms in Eq.~(\ref{total2}). The
tunneling matrix element in Eq.~(\ref{2n}) is determined using the
particle-conserving Bogoliubov transformation and is dependent
only on $p$ and $k_1$. Therefore, by doing the sum over the other
momenta $k_j$, one gets the following result:
\begin{eqnarray}\label{Gamma02}
\Gamma_{2}^{\rm{op}}&=&\pi\!\sum_{p,k_1}{|\bra{\!N+1,p}H_{T}\ket{N+2,k_1}|^2
\delta(E_{p}\!-E_{k_1}\!-\omega_{+})}  \nonumber \\
&\times&
\exp\left(-\frac{E_{k_1}}{T}\right)\sum\limits_{n}\frac{\left[z_b(T,\delta_b)
\right]^{2n-1}}{(2n-1)!Z_{\rm{ev}}}.
\end{eqnarray}
Changing the sum to an integral and integrating over $E_p$, we
obtained the following expression:
\begin{eqnarray}
\label{Gamma2}
\Gamma_{2}^{\rm{op}}\!&\!=\!&\!\frac{g}{4\pi}\int_{\Delta_b}^{\infty}{\!dE_k}\frac
{(E_k(E_k\!+\omega_{+})\!-\Delta_r\Delta_b)}{\sqrt{((E_k+\omega_{+})^2-\Delta_r^2)(E_k^2-\Delta_b^2)}} \nonumber\\
 \nonumber\\
\!&\!\times\!&\!\Theta(\!E_k\!+\!\omega_{+}\!-\!\Delta_r)\!\exp{\left(\!-\frac{E_k}{T}\right)}
\!\tanh\left(\!z_{b}(T,\!\delta_{b}\!)\right).
\end{eqnarray}
The integration can be performed assuming that mismatch between
superconducting gap energies in the box and reservoir is small,
and $\Delta_b-\Delta_r+\omega_{+}> 0$. Comparing
Eqs.~(\ref{Gamma11}) and ~(\ref{Gamma2}), one notices that the
answer for $\Gamma_{2}^{\rm{op}}$ can be expressed via
$\Gamma_{1}^{\rm{op}}$ by permuting $\Delta_r \leftrightarrow
\Delta_b$
\begin{eqnarray}\label{G2}
\Gamma_{2}^{\rm{op}}(\omega_{+},\Delta_r,\Delta_b)\!=\!\Gamma_{1}^{\rm{op}}(\!\omega_{+},\!\Delta_b,\!\Delta_r)\!\tanh\left(z_{b}(T,\delta_{b})\right).
\end{eqnarray}

Taking into account  Eq.~(\ref{Gamma0}) and Eq.~(\ref{G2}), we
find the total quasiparticle decay rate for the open system
\begin{eqnarray}
\Gamma_{\ket{N+1} \leftarrow \ket{+}}^{\rm{op}}&=& W(\omega_{+},
\Delta_r,\Delta_b)+W(\omega_{+},\Delta_b,\Delta_r) \nonumber \\
&\times& \tanh\left(z_{b}(T,\delta_{b})\right),
\end{eqnarray}
where the first term corresponds to the first mechanism and is
dominating for the open system. The simplified results for
$\Gamma_{\ket{N+1} \leftarrow \ket{+}}^{\rm{op}}$ are discussed in
Sec. IV.

Decay rate evaluation for the ground state of the qubit $\ket{-}$
can be done similarly provided that the final state $\ket{N+1}$ of
the qubit is lower in energy than the initial state $\ket{-}$.

\section{States of the qubit in an isolated system}
\subsection{Thermodynamic properties of the qubit with fixed
number of electrons}

When the number of electrons in the qubit is fixed, parity effects
become important at low temperatures, $T<T_{r}^*,T_{b}^*$, where
$T_{b}^*$ is defined in Eq.~(\ref{tstar}) and $T_{r}$ is equal to
\begin{equation}\label{tstar_r}
T_{r}^*=\frac{\Delta_r}{\ln(\Delta_r/\delta_r)}.
\end{equation}
Here, $\delta_r$ is the mean level spacing in the reservoir. The
density of quasiparticles in the qubit with even number of
electrons is small, $n \propto \exp\left(-2\Delta/T\right)$, since
at low temperatures all electrons are paired, and it costs energy
$2\Delta$ to break a Cooper pair. In the odd-charge state, an
unpaired electron is present in the system even at zero
temperature. It is important to understand the probability of
finding a quasiparticle in the CPB since the qubit will not work
if an extra electron is present in the box. To find out whether it
is favorable or not for a quasiparticle to reside in the CPB, we
calculated the difference in free energy $\Delta F=F_1-F_0 $
between two states: with  and without a quasiparticle in the box
($F_1$ and $F_0$, respectively). At the operating point, the free
energy difference for the qubit with even ($\Delta F_{\rm{ev}}$)
and odd ($\Delta F_{\rm{odd}}$) total number of electrons is given
by the following expressions:
\begin{eqnarray}
 \Delta F_{\rm{ev}} &=& -E_c+\frac{E_J}{2}-T\ln[\tanh(z_b(T,\delta_{b}))] \nonumber\\
&-&T\ln[\tanh(z_r(T,\delta_{r}))]
\end{eqnarray}
\begin{eqnarray}
 \Delta F_{\rm{odd}} &=& -E_c+\frac{E_J}{2}-T\ln[\tanh(z_b(T,\delta_{b}))] \nonumber\\
&-&T\ln[\coth(z_r(T,\delta_{r}))]
\end{eqnarray}
Negative value of $\Delta F$ indicates that free energy is lower
for a quasiparticle in the CPB. Using these expressions, we can
calculate thermodynamic probability $P(T)$ to find an unpaired
electron in the box as a function of the physical parameters
\begin{equation}
P(T)=\frac{Z_1}{Z_1+Z_0}=\frac{1}{\exp\left(\frac{\Delta
F}{T}\right)+1},
\end{equation}
where $Z_{1(0)}=\exp\left(-\beta F_{1(0)}\right)$ is the partition
function with one (zero) unpaired electrons in the box. The
expression for the free energy difference $\Delta F_{op}$ for the
open system (fixed chemical potential regime) can be obtained
using $ \Delta F_{\rm{ev}}$ and taking the limit of infinite
volume of the reservoir ($\delta_{r}\rightarrow 0$). Temperature
dependence of the quasiparticle probability in the CPB for
realistic parameters is plotted in Fig.2.
\begin{figure}
\centering
\includegraphics[width=3.4in]{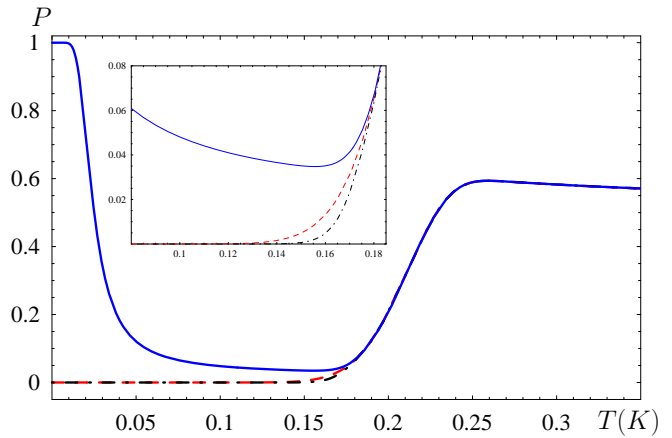}
\caption{(color online). Main panel: temperature dependence of the
number of quasiparticles in the box at the operating point
($N_g=1$). Dash-dot (black) line corresponds to even number of
electrons, solid (blue) line - odd number of electrons, dash (red)
line - open system. Physical parameters are chosen in correspondence
to typical qubit experiments: $\Delta_r=\Delta_b$ = 2.4K, $E_c$ =
0.25K, $E_J$ = 0.3K, $T^*_b$ = 210mK, and $T^*_r$ = 160mK (see
Eqs.~(\ref{tstar}) and~(\ref{tstar_r}) for definition of $T^*_b$ and
$T^*_r$). Inset: temperature dependence of the number of
quasiparticles in the CPB in the vicinity of $T_r^*$. }
\end{figure}

As shown in Fig.2, at high temperatures $T>T_{b}^*$ the
probability of having an extra electron in the CPB coincides for
an open and isolated qubit. At this temperature the number of
thermal quasiparticles in the system is large, and parity effects
are not important. Parity effects start to manifest themselves
below the characteristic temperature $T_b^*$, when the number of
quasiparticles in the box is of the order of unity. As can be seen
from Fig.2, at the temperature $T_r^*$ the probability of having a
quasiparticle in the CPB is negligible in the even-charge state,
as well as in the open system. In the case of odd charge state of
a qubit, lowering the temperature enhances quasiparticle
poisoning~\cite{my}. This effect can be explained as a competition
of two contributions to the free energy: a charging energy gained
by tunneling to the box and entropy contribution proportional to
the ratio of the volumes of the reservoir and box, $\sim V_r/V_b$.
At the temperature $T_s$
\begin{equation}
T_s \simeq \frac{E_c-E_J/2}{\ln\left(\frac{V_r}{V_b}\right)}
\end{equation}
the entropy contribution becomes smaller, and the quasiparticle
resides in the CPB. Thus, for the odd-charge state there is only a
certain intermediate temperature range when qubit can work,
\textit{i.e.} can be prepared in the ``good" quantum state. For
physical parameters used in Fig.2 $T_s$ is approximately equal to
20mK.

The presence of a quasiparticle in the box can be studied
experimentally by measuring the periodicity of Coulomb staircase
~\cite{Aumentado, Turek, Guillaume, Lehnert, Mannik, Bouchiat}.
According to the obtained results, the Coulomb staircase for an
open system should be $2e$ periodic below the temperature
$\widetilde{T}_{b}^*$. The qubit with fixed number of electrons
should have two distinct types of behavior corresponding to even
and odd total number of electrons in the box and reservoir. In the
former case the Coulomb staircase is similar to the one in the
open system, while in the latter Coulomb staircase is $1e$
periodic for temperatures above $T_{b}^*$, then $2e$ periodic from
$T_{b}^*$ to $T_s$, and again $1e$ periodic for $T< T_{s}$.

In principle the probability to find a quasiparticle in the CPB
can be lowered, and the qubit can be brought into the desired
quantum state. We discuss several ways of doing this in Sec. IV.
However, even if the quasiparticle is in the reservoir at the
initial moment of time, once the qubit is excited and quantum
oscillations emerge, the time of the oscillations is determined by
the kinetics, \emph{i.e.}, by quasiparticle tunneling rate.

\subsection{Quasiparticle decay rate in the isolated system}

Let us turn to the discussion of the lifetime of the charge qubit
in the regime with the fixed number of electrons. In order to
calculate the quasiparticle decay rate, we proceed in the same
manner as in the open system. The decay rate for the even number
of electrons is calculated by averaging over initial states with
even parity density matrix $\rho_{2n}(\beta H_0)$. This situation
corresponds to having an even number of electrons in the box and
reservoir. The appearance of quasiparticles in the system occurs
at the expense of breaking Cooper pairs. Using the results of
analogous calculation in Eq.~(\ref{G2}), we can write the
expression for the total decay rate
\begin{eqnarray}
\Gamma_{\ket{N+1} \leftarrow \ket{+}}^{\rm{ev}}&=&W(\omega_{+},
\Delta_r,\Delta_b)\tanh\left(z_{r}(T,\delta_{r})\right) \nonumber\\
&+&W(\omega_{+}, \Delta_b,
\Delta_r)\tanh\left(z_{b}(T,\delta_{b})\right),
\end{eqnarray}
where $W(\omega_{+}, \Delta_r,\Delta_b)$ is defined in
Eq.~(\ref{Gamma0}). The first term here corresponds to the first
mechanism given by Eq.~(\ref{total1}) and averaged over the
even-parity initial state.

In the odd-charge case the decoherence rate is the largest since a
quasiparticle is present in the system at $T=0$. Initial
configuration of the system corresponds to having an odd number of
quasiparticles in the reservoir. The reduced density matrix for
this initial state $\rho_{2n-1}(\beta H_0)$ is then given by
\begin{equation}
\rho_{2n-1}(\beta H_0)=Tr_{\{k\}}\rho(\beta
H_0)=\frac{\exp\left(-\sum\limits_{j=1}^{2n-1}{\frac{E_{p_j}}{T}}\right)}{(2n-1)!
Z_{\rm{odd}}},
\end{equation}
where $Z_{\rm{odd}}=\sinh\left(z_{r}(T,\delta_{r})\right)$. Using
Eq.~(\ref{total1}), we write the contribution to the decay rate of
the first mechanism
\begin{eqnarray}
\Gamma_{1}^{\rm{odd}}&=&
2\pi\sum_{n,p_j,k}|\bra{N+1,k,\{p\}_{2n-2}}H_{T}\ket{+,\{p\}_{2n-1}}|^2
\nonumber \\
&\times& \delta(E_{k}-E_{p_1}-\omega_{+})(2n-1)\rho_{2n-1}(\beta
H_0).
\end{eqnarray}
Going through the same arguments as in Eq.~(\ref{Gamma02}),  the
expression for $\Gamma_{1}^{\rm{odd}}$ can be simplified
\begin{eqnarray}
\Gamma_{1}^{\rm{odd}}&=&\!\pi\sum_{p_1,k}{|\bra{N+1,k}H_{T}\ket{N,p_{1}}|^2\delta(E_{k}-E_{p_1}-\omega_{+})} \nonumber \\
\nonumber \\
&\times&\exp\left(-\frac{E_{p_1}}{T}\right)\sum\limits_{n=1}^{\infty}\frac{\left[z_r(T,\delta_{r})\right]^{2n-2}}{(2n-2)!Z_{\rm{odd}}}
\end{eqnarray}
Summing over $E_k$, we get
\begin{eqnarray}\label{2}
\!\Gamma_1^{\rm{odd}}\!&\!=\!&\!\frac{g}{4\pi}\int_{\Delta_r}^{\infty}{\!dE_p}\frac
{(E_p(E_p+\omega_{+})-\Delta_r
\Delta_b)}{\sqrt{((E_p+\omega_{+})^2-\Delta_b^2)(E_p^2-\Delta_r^2)}} \nonumber\\
& &\nonumber\\
\!&\!\times\!&\!\Theta(E_p\!+\!\omega_{+}\!-\!\Delta_b)\exp\!\left(\!-\frac{E_p}{T}\right)\!\coth\left(z_{r}(T,\!\delta_{r})\right)
\end{eqnarray}
Taking into account Eq.~(\ref{Gamma11}) and Eq.~(\ref{Gamma0}),
$\Gamma_1^{\rm{odd}}$ is equal to
\begin{equation}
\Gamma_1^{\rm{odd}}=W(\omega_{+},\Delta_r,\Delta_b)\coth\left(\!z_{r}\!(T,\!\delta_{r})\right).
\end{equation}

In order to find the contribution of the second mechanism, one has
to average over initial states of the CPB. The initial
configuration of the box corresponds to the even-charge state and
is the same for open and isolated qubits. Therefore, contribution
of the second mechanism, $\Gamma_2^{\rm{odd}}$, is given by
Eq.~(\ref{G2}).

Total decay rate $\Gamma_{\ket{N+1} \leftarrow
\ket{+}}^{\rm{odd}}$ with odd number of electrons in the qubit is
the sum of $\Gamma_1^{\rm{odd}}$ and $\Gamma_2^{\rm{odd}}$
\begin{eqnarray}
\Gamma_{\ket{N+1}\! \leftarrow \ket{+}}^{\rm{odd}}&=&W(\omega_{+},
\Delta_r,\Delta_b)\coth\left(z_{r}(T,\delta_{r})\right)\nonumber\\
&+&W(\omega_{+},\Delta_b,\Delta_r)\tanh\left(z_{b}(T,\delta_{b})\right)
\end{eqnarray}
At low temperatures, $T<T_{r}^*,T_{b}^*$, the first term here is
dominant as $z(T,\delta)\ll 1$. The detailed analysis of the low
temperature asymtotics for different experimental regimes is
presented below.

\section{Discussion of the Results}

Temperature dependence of the quasiparticle decay rate for
different experimental realizations of the qubit is shown in
Fig.3. As it is clear from the figure, at experimentally relevant
temperatures $T \ll T_{r}^*,T_{b}^*$, the largest decay rate
corresponds to the odd-charge case. In the vicinity of $T^*_r$
defined in Eq.~(\ref{tstar_r}), the decay rate is growing quickly
due to the appearance of a large number of quasiparticles in the
reservoir. As we approach the temperature $T=T_{b}^*$ given in
Eq.~(\ref{tstar}), which corresponds to the appearance of
quasiparticles in the Cooper pair box, parity effects become
irrelevant and decoherence rates for different cases coincide.
\begin{figure}[!htb]
\centering
\includegraphics[width=3.3in]{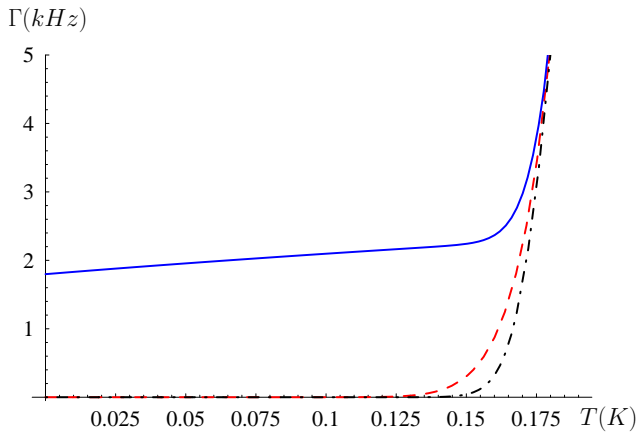}
\caption{(color online) Temperature dependence of the
quasiparticle decay rate. Dash (red) line corresponds to the open
system, dash-dot (black) - even number of electrons, solid (blue)
- odd number of electrons. Here, we used the same physical
parameters as specified in Fig. 2.}
\end{figure}

Results obtained in Sec. II and III allow us to quantitatively
estimate the decoherence rate due to the presence of
quasiparticles in the system. For simplicity we assume that
superconducting gap energies are the same in the box and
reservoir, $\Delta_b=\Delta_r=\Delta$, and temperature is low,
$T<T_{r}^*,T_{b}^*\ll \omega_{+}$, corresponding to typical qubit
experiments ($\omega_{+}$ is defined in Eq.~(\ref{omega})). In
this approximation, for an ``open" qubit the decay rate is
\begin{eqnarray}\label{as1}
\Gamma_{\ket{N+1} \leftarrow \ket{+}}^{\rm{op}} \!&\!\simeq\!&\!
\frac{
g\sqrt{T\Delta}}{8\sqrt{\pi}}\sqrt{\frac{\omega_{+}}{\Delta+\omega_{+}/2}}\exp\left(-\frac{\Delta}{T}\right) \nonumber\\
& &\nonumber\\
\!&\!+\!&\!\frac{gT}{8\sqrt{\pi}}\sqrt{\frac{\omega_{+}}{\Delta\!+\!\omega_{+}/2}}\frac{\Delta}{\delta_b}\exp\left(\!-\frac{2\Delta}{T}\right)\!.
\end{eqnarray}
The leading contribution to the decay rate for the open system
$\Gamma_{\ket{N+1} \leftarrow \ket{+}}^{\rm{op}}$ is proportional
to $\exp(-\Delta/T)$ since energy equal to $\Delta$ is required to
bring an electron from the normal parts.

In the even-charge case the leading contribution to
$\Gamma_{\ket{N+1} \leftarrow \ket{+}}^{\rm{ev}}$ is exponentially
small, $\propto\exp(-2\Delta/T)$, in accordance with the energy
necessary to break a Cooper pair,
\begin{eqnarray}\label{as2}
\Gamma_{\ket{N+1} \leftarrow \ket{+}}^{\rm{ev}} \!&\!\simeq\!&\!
\frac{g T}{8\sqrt{\pi}}
\sqrt{\frac{\omega_{+}}{\Delta\!+\!\omega_{+}/2}}\frac{\Delta}{\delta_r}\exp\left(\!-\frac{2\Delta}{T}\right) \nonumber\\
& &\nonumber\\
\!&\!+\!&\!\frac{gT}{8\sqrt{\pi}}\sqrt{\frac{\omega_{+}}{\Delta\!+\!\omega_{+}/2}}\frac{\Delta}{\delta_b}\exp\left(\!-\frac{2\Delta}{T}\right)\!.
\end{eqnarray}
Note that Eq.~(\ref{as2}) and the last term in Eq.~(\ref{as1}) are
essentially the upper bounds for the contributions to the decay
rate coming from unpaired electrons which originate in the
isolated qubit and in the Cooper pair box, respectively. Indeed,
we assumed in the derivation of Eqs.~(\ref{as1}) and~(\ref{as2})
that the decay rate is limited by the thermodynamic probability of
the unpaired state, without discussing the kinetics of
pair-breaking leading to such state. It is clear from Fig. 3,
however, that the above mentioned contributions are negligibly
small at low temperatures.

The decay rate of the qubit with an odd number of electrons
$\Gamma_{\ket{N+1} \leftarrow \ket{+}}^{\rm{odd}}$ is much larger
\begin{eqnarray}\label{as3}
\Gamma_{\ket{N+1} \leftarrow \ket{+}}^{\rm{odd}} \!&\!\simeq\!&\!
\frac{g\delta_r}{8\sqrt{\pi}}\sqrt{\frac{\omega_{+}}{\Delta+\omega_{+}/2}}\nonumber\\
& &\nonumber\\
\!&\!+\!&\!\frac{g T}{8\sqrt{\pi}}
\sqrt{\frac{\omega_{+}}{\Delta\!+\!\omega_{+}/2}}
\frac{\Delta}{\delta_b}\!\exp\left(\!-\frac{2\Delta}{T}\right)\!.
\end{eqnarray}
The leading contribution to the decay rate is temperature
independent since the number of quasiparticles is finite even at
$T=0$. According to the Eq.~(\ref{as3}), the lifetime of qubit in
the odd-charge case is determined by the conductance of the tunnel
junctions and mean level spacing of the reservoir. For typical
experiments quantum conductance, $g$ is of the order of one;
$\delta_r$ depends on the volume of the reservoir and varies
between $10^{-10} \mbox{ and } 10^{-12}$ eV. With these
parameters, decay rate $\Gamma_{\ket{N+1} \leftarrow
\ket{+}}^{\rm{odd}}$ can be estimated as $10^5-10^3\mbox{ Hz}$
consequently. This is a substantial contribution to the
decoherence of the isolated charge qubits, which limits qubit
operation on a fundamental level. However, this decay rate is much
smaller than the present estimates for decoherence in charge
qubits; see, for example, the recent review by Devoret \emph{et.
al}~\cite{Devoret}.

We assumed so far that in the initial state an unpaired electron
resides in the reservoir and finally (after the relaxation) ends up
in the box. Then, one can tune the qubit to the charge degeneracy
between 1$e$ and 3$e$ and still have charge oscillations for some
time until the quasiparticle escapes into the reservoir. In this
case, the quasiparticle escape rate can be calculated in a similar
way and is proportional to the conductance of the tunnel junction
and level spacing in the Cooper-pair box: $\Gamma^{\rm{odd}}\propto
g \delta_b$.

In principle, quasiparticle poisoning can be decreased by tuning
the proper parameters of the system such as superconducting gap
energies $\Delta_{r,b}$, charging and Josephson energies,
temperature, volumes of the box and reservoir. For example, it can
be done by adjusting gap energies $\Delta_{r,b}$ with the help of
oxygen doping~\cite{Aumentado} or magnetic field~\cite{Turek}. The
latter is easy to implement since magnetic field is already used
in charge qubits to tune Josephson energy $E_J$. In addition to
the suppression of gap energies, large magnetic field ($H>
H_{c1}$) can create vortices in the reservoir. The vortex acts as
a quasiparticle trap since an unpaired electron gains gap energy
$\Delta$ by residing inside the vortex.

\section{Conclusions}

We demonstrated that the presence of quasiparticles in the
superconducting charge qubit leads to the decay of quantum
oscillations. Two experimental realizations of charge qubit are
considered here, corresponding to open and isolated system (in the
former, the number of electrons is not fixed). Once the qubit is
excited and quantum oscillations emerge, the decay of these
oscillations is determined by the quasiparticle tunneling rate to
the Cooper pair box. We calculated temperature dependence of the
quasiparticle decay rate in the charge qubit. This decay rate is
exponentially suppressed in the open system as well as in the
isolated system with an even number of electrons. However, in the
case with an odd number of electrons in the system, the
quasiparticle decay rate is not exponentially suppressed and is
estimated to be $10^5-10^3$ Hz depending on the volume of the
superconducting reservoir and conductance of the tunnel junctions.

\begin{acknowledgments}

We thank R. Schoelkopf and A. Wallraff for useful discussions. This
work was supported by NSF grants DMR 02-37296, DMR 04-39026 and EIA
02-10736.

\end{acknowledgments}


\thebibliography{plain}

\bibitem{Nakamura} Y. Nakamura, Y. A. Pashkin, and J. S. Tsai, Nature {\bf 398}, 786
(1999).
\bibitem{Wallraff} A. Wallraff, D. I. Schuster, A. Blais, L. Frunzio, R.-S. Huang, J. Majer, S. Kumar, S. M. Girvin, and R. J. Schoelkopf, Nature (London) {\bf 431}, 162 (2004).
\bibitem{Chiorescu}  I. Chiorescu, Y. Nakamura, C. J. P. M. Harmans, and J. E. Mooij, Science {\bf 299}, 1869 (2003).
\bibitem{Vion} D. Vion, A. Aassime, A. Cottet, P. Joyez, H. Pothier, C. Urbina, D. Esteve, and M.H. Devoret, Science {\bf 296}, 286 (2002).
\bibitem{Blais}  A. Blais, R.-S. Huang, A. Wallraff, S. M. Girvin, and R. J. Schoelkopf, Phys. Rev. A {\bf 69}, 062320 (2004)
\bibitem{Devoret}  M. H. Devoret, A. Wallraff, J. M. Martinis,
cond-mat/0411174.
\bibitem{Tinkham} M. T. Tuominen, J. M. Hergenrother, T. S. Tighe, and M. Tinkham, Phys. Rev. Lett. {\bf 69}, 1997
(1992).
\bibitem{Hekking} F. W. J. Hekking, L.I. Glazman, K. A. Matveev, and R. I. Shekhter, Phys. Rev. Lett. {\bf 70}, 4138
(1993).
\bibitem{Lafarge} P. Lafarge, P. Joyez, D. Esteve, C. Urbina, and M.H. Devoret, Phys. Rev. Lett. {\bf 70}, 994
(1993).
\bibitem{Shnirman} Y. Makhlin, G. Schoen, and A. Shnirman, Rev. Mod. Phys. {\bf 73}, 357
(2001).
\bibitem{Turek} B. Turek, J. Majer, A. Clerk, S. Girvin, A. Wallraff, K. Bladh, D. Gunnarson, T. Duty, P. Delsing and R.
Schoelkopf, Proceedings of Applied Superconductivity Conference,
Jacksonville, FL 2004.
\bibitem{Guillaume} A. G. Guillaume, J. F. Schneiderman, P. Delsing, H. M. Bozler, and
P. M. Echternach, Phys. Rev. B {\bf 69}, 132504 (2004).
\bibitem{Lehnert} K. W. Lehnert, K. Bladh, L. F. Spietz, D. Gunnarson, D. I. Schuster,
P. Delsing, and R. J. Schoelkopf, Phys. Rev. Lett. {\bf 90},
027002 (2003).
\bibitem{Aumentado} J. Aumentado, M. W. Keller, J. M. Martinis, M. H. Devoret, Phys. Rev. Lett.
{\bf 92}, 66802 (2004).
\bibitem{Mannik} J. M¨annik and J. E. Lukens, Phys. Rev. Lett. {\bf 92}, 057004 (2004).
\bibitem{Schrieffer} J.R. Schrieffer, \textit{Theory of Superconductivity}, (Oxford : Advanced Book Program, Perseus, 1999).
\bibitem{Bouchiat} V. Bouchiat, D. Vion, P. Joyez, D. Esteve, and M. H. Devoret, Physica Scripta, {\bf T76},
165 (1998).
\bibitem{Matveev} K.A. Matveev, L.I. Glazman, and R.I. Shekhter
Mod. Phys. Lett. B {\bf 8}, 1007 (1994).
\bibitem{my} This effect was observed experimentally; see
ref.[14] and references therein.
\end{document}